%% file: a1835.tex
\documentclass[letters, usenatbib]{mn2e}

\pdfoutput=1

\usepackage{url}
\usepackage{fixltx2e}
\usepackage{mathptmx}
\usepackage[pdftex]{graphicx}

\include{defn}

\begin{document}
\title[Turbulence in the core of Abell~1835]
{A direct limit on the turbulent velocity of the intracluster medium
  in the core of Abell 1835 from \emph{XMM-Newton}}

\author
[J.S. Sanders et al]
{J.~S. Sanders$^1$,
  A.~C. Fabian$^1$, R.~K. Smith$^2$ and J.~R. Peterson$^3$
  \\
  $^1$ Institute of Astronomy, Madingley Road, Cambridge. CB3 0HA\\
$^2$ MS 6, 60 Garden Street, Harvard-Smithsonian Center
for Astrophysics, Cambridge, MA 02138, USA \\
$^3$ Department of Physics, Purdue University, 525 Northwestern
Avenue, West Lafayette, IN 47907-2036, USA
}
\maketitle

\begin{abstract}
  We examine deep \emph{XMM-Newton} Reflection Grating Spectrometer
  (RGS) observations of the X-ray luminous galaxy cluster A1835. For
  the first time in a galaxy cluster we place direct limits on
  turbulent broadening of the emission lines. This is possible because
  the coolest X-ray emitting gas in the cluster, which is responsible
  for the lines, occupies a small region within the core.  The most
  conservative determination of the 90 per cent upper limit on
  line-of-sight, non-thermal, velocity broadening is $274\kmps$,
  measured from the emission lines originating within 30~kpc
  radius. The ratio of turbulent to thermal energy density in the core
  is therefore less than 13 per cent.  There are no emission lines in
  the spectrum showing evidence for gas below $\sim 3.5$~keV. We
  examine the quantity of gas as a function of temperature and place a
  limit of $140 \Msunpyr$ (90 per cent) for gas cooling radiatively
  below 3.85~keV.
\end{abstract}


\section{Introduction}
Little direct information is known about the dynamical state or
microphysics of the intracluster medium (ICM). Simulations of galaxy
clusters within a cosmological environment predict that as matter
accretes along filaments, or as clusters merge, turbulent motions
should be set up in the ICM. This energy should then cascade from
large to small scales where it can be dissipated. Simulations have
predicted that even in the most relaxed clusters there are substantial
flows of ICM (e.g. \citealt{Evrard90}). In these simulations the
fraction of pressure support in gas motions is typically 5 to 15 per
cent. Most simulations, however, have not included physical processes
which may be relevant, such as viscosity or magnetic fields.

In addition, it is observed that the active galactic nuclei (AGN) in
cluster cores inject jets and bubbles of relativistic plasma into
their surroundings. The feedback of energy into the cluster is thought
to combat the high radiative losses of energy which would otherwise
lead to substantial cooling rates (for reviews see
\citealt{PetersonFabian06} and \citealt{McNamaraNulsen07}).  It is
predicted that feedback could induce gas motions in the core in the
range $500-1000\kmps$ \citep{Bruggen05}. Some models also require
turbulence in order to heat the ICM using the energy in inflated
cavities (e.g. \citealt{Scannapieco08}). Strong turbulent motions
would provide significant non-thermal pressure support. This would
bias the determination of cluster mass profiles measured under the
assumption of hydrostatic equilibrium. The measurement of turbulence
is thus important for mass determination, as well as AGN feedback.

Evidence for resonance scattering has also been used to place upper
limits on turbulence in elliptical galaxies. In NGC~4636, \cite{Xu02}
examined the strength of the resonantly scattered 15{\AA}
Fe~\textsc{xvii} line relative to the line at {17.1\AA}, inferring
scattering. As turbulence broadens emission lines and prevents
scattering, they obtained an upper limit on the turbulent velocity
dispersion of 10 per cent of the sound speed. For the same object,
\cite{Werner09} calculated a limit of 25 per cent of the sound speed,
with an implied maximum of 5 per cent of energy in turbulent motions.

The Perseus cluster is an interesting test case. The tens of
  kpc long quasi-linear filaments \citep{FabianPerFilament03} imply
laminar flow, significant viscosity in the ICM and low levels of
turbulence on the scale of the filaments.  However, the lack of
evidence for resonant scattering in the strongest 6.7~keV line of
He-like iron in Perseus argues for velocity spread of at
least half the sound velocity \citep{Churazov04}.
\cite{RebuscoDiff05} also require motions of $300\kmps$ on scales of
$20\kpc$ to account for the overall metallicity profile in Perseus.

Comparing the gravitational mass profiles derived from optical and
X-ray data in M87 and NGC~1399, \cite{Churazov08} obtained combined
nonthermal pressures of $\sim 10$ per cent of the thermal gas
pressure.  \cite{Reynolds05} simulated the evolution of cavities in
the ICM, requiring viscous processes in order for the cavities to
retain the observed spherical cavity shapes and prevent them from
being shredded by instabilities \citep{Reynolds05,Dong09}.  A
\emph{lower} limit of 10 per cent of the total pressure in turbulence
was found in the Coma cluster by \cite{Schuecker04}. The Coma cluster
is unrelaxed with two central galaxies. It is unlikely that it can be
directly compared to relaxed clusters or ellipticals. In summary, the
evidence for turbulence in relaxed clusters is uncertain.

A1835 (at $z=0.2523$) is the most luminous cluster in the BCS sample
\citep{Ebeling98}, with an inferred cooling rate of $\sim 1000
\Msunpyr$ in the absence of heating \citep{Allen96}.  The
optical-spectroscopic-derived rate of star formation in the central
galaxy is 40--70\Msunpyr \citep{Crawford99}. The IR luminosity of the
object is $\sim 7 \times 10^{11} \Lsun$ \citep{Egami06}, implying a
star formation rate of $\sim 125 \Msunpyr$. In addition there is
$\sim10^{11} \Msun$ of molecular gas in the cluster \citep{Edge01}.

Despite the evidence for star formation, cool gas and the high X-ray
luminosity, little X-ray emitting gas was seen in this cluster below
1--2~keV when it was observed using the \emph{XMM-Newton} RGS
  instruments \citep{Peterson01}. Less than $200\Msunpyr$ cools below
  2.7~keV (90 per cent confidence).  AGN feedback, at a level of
  $1.6\times 10^{45} \ergps$, prevents most of the X-ray gas from
  cooling.  Later observations confirmed the lack of X-ray cool gas is
  a common feature of clusters (e.g. \citealt{Peterson03}).

In this paper we examine deep RGS observations of A1835 totalling
254~ks, placing limits on velocities in the gas and the amount of cool
gas detectable in X-rays. We use a Hubble constant of $70 \kmpspMpc$
and the relative Solar metallicities of \cite{AndersGrevesse89}. 1
arcsec corresponds to 3.94~kpc at $z=0.2523$. Errors quoted are
$1\sigma$ unless stated otherwise.

\section{RGS spectra}

We examined three \emph{XMM-Newton} RGS observations (0098010101,
0551830101 and 0551830201).  We extracted the RGS spectrum from the
datasets using the standard \textsc{sas} 9.0 \textsc{rgsproc} tool. We
used a 90 per cent PSF width extraction region (corresponding to a
half-width of approximately 30~arcsec), and a 90 per cent pulse-height
distribution energy selection. To eliminate flares, we removed time
periods with a count rate greater than $1\ps$ on CCD 9, selecting
events with FLAG set to 8 or 16 and a cross dispersion angle greater
than $1.5 \times 10^{-4}$.

Background spectra were generated from the observations, extracted
from the region beyond a 98 per cent of PSF width. We tested that the
backgrounds were essentially source free by comparison with template
background spectra.  The wavelength binning option was used so that we
could combine the spectra from both RGS instruments using
\textsc{rgscombine}. We combined the spectra, backgrounds and
responses from all observations, treating the first and second order
spectra separately.

\begin{table}
  \caption{Single temperature RGS spectral fitting results. The
    results with \textsc{rgsxsrc} use that model to account for the
    spatial extent of the source.}
  \begin{tabular}{lcc}
    \hline
    Parameter                & Without \textsc{rgsxsrc} & With
    \textsc{rgsxsrc} \\ \hline
    kT (keV)                 & $3.67 \pm 0.16$    & $3.71 \pm 0.16$ \\
    O  (\Zsun)               & $0.17 \pm 0.03$    & $0.21 \pm 0.03$ \\
    Ne (\Zsun)               & $0.25 \pm 0.04$    & $0.30 \pm 0.05$ \\
    Mg (\Zsun)               & $0.17 \pm 0.08$    & $0.20 \pm 0.09$ \\
    Si (\Zsun)               & $0.19 \pm 0.06$    & $0.19 \pm 0.06$ \\
    Fe (\Zsun)               & $0.18 \pm 0.02$    & $0.20 \pm 0.02$ \\
    normalization (cm$^{-5}$)& $0.0104 \pm 0.0001$& $0.0107 \pm
    0.0001$ \\
    C statistic              & 3947.9             & 3956.4 \\
    \hline
  \end{tabular}
  \label{tab:1t}
\end{table}

\begin{figure*}
  \includegraphics[width=\textwidth]{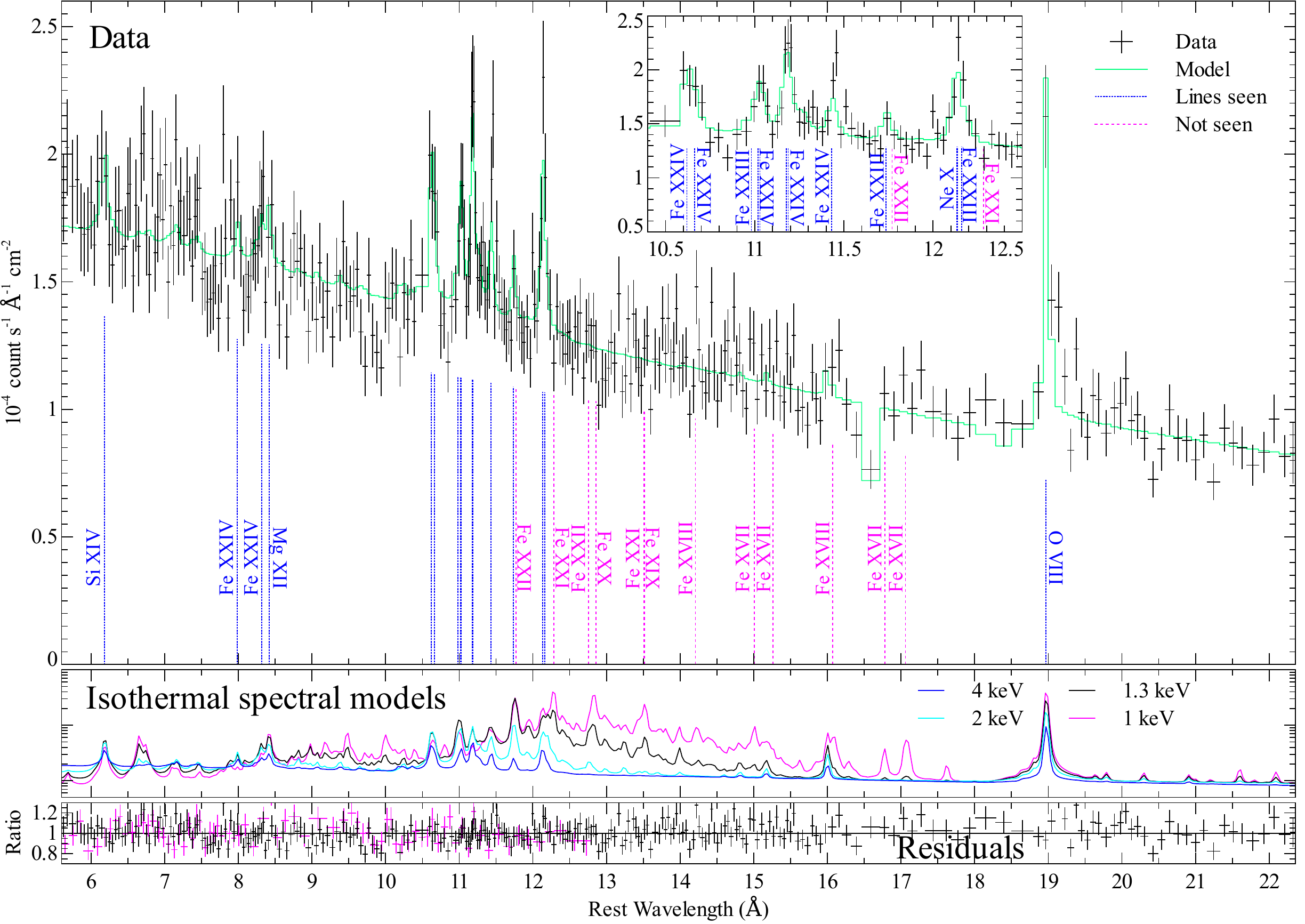}
  \caption{(Top panel). Combined RGS 1 and 2 spectrum from the 90
    per~cent extraction region and best fitting single-temperature
    model. The data have been rebinned to have a signal to noise ratio
    of 10 and divided by the effective area as function of
    wavelength. The spectral feature close to 16{\AA} seen in the
    spectral model next to Fe~\textsc{xviii} is O~\textsc{viii}.
    (Centre panel) Isothermal Solar metallicity spectral models at
    different temperatures smoothed with the RGS response. (Bottom
    panel) Rebinned residuals of the model fit to the data. The
    magenta points at short wavelength are the residuals from the
    second order spectrum.}
  \label{fig:spectrum}
\end{figure*}

\subsection{Single temperature model}
\label{sect:1t}
We show in Fig.~\ref{fig:spectrum} the first order combined spectra
with the best fitting thermal model. We used the \textsc{spex} version
2.00.11 thermal model \citep{Kaastra00}, converted to an \textsc{apec}
format table for \textsc{xspec}, with Galactic absorption fixed at
$2.04\times 10^{20} \pcmsq$ \citep{Kalberla05}. We fitted both
spectral orders simultaneously, using a wavelength range of 7 to
28{\AA} and 7 to 16{\AA} for the 1st and 2nd orders, respectively.
The O, Ne, Mg, Si and Fe metallicities were allowed to be free in the
fit with the other elemental abundances tied to Fe, except He which
was fixed to Solar. Ni and Ca lines in the fitted
  spectral region are too weak to fit separately. We allow a
normalization offset between the two orders (the 2nd order gives a
best fitting normalization 3 per cent greater than the 1st order
spectrum).  We fit the spectrum minimizing the C statistic on the
ungrouped spectra in \textsc{xspec} 12.5.1 \citep{ArnaudXspec}.  The
thermal broadening option was enabled when fitting the spectra.

The C statistic does not provide a goodness of fit, but the
\textsc{xspec} goodness command calculated 50.9 per cent of Monte
Carlo realisations of the best fitting model have a better fit
statistic than the real data, indicating a good fit. If the spectra
are fit with the same model but using $\chi^2$ statistics, the reduced
$\chi^2$ is $1.02 = 2253/2218$ (grouping to have at least 20 counts
per channel). Shown in the lower panel of Fig.~\ref{fig:spectrum} are
the rebinned residuals of the single temperature model to the first
and second order spectra. The best fitting model parameters are shown
in Table~\ref{tab:1t} (under the column without \textsc{rgsxsrc}). The
low Fe abundance is interesting. This could be significantly affected
by how much continuum emission is from hotter plasma ($>5$~keV)
outside the cluster core, which is not well constrained by lines in
the RGS band. We note that the line complexes at 11.4 and 12.2{\AA}
are underpredicted by our model, which we could not match by adding
additional temperature components.

\begin{figure}
  \includegraphics[width=\columnwidth]{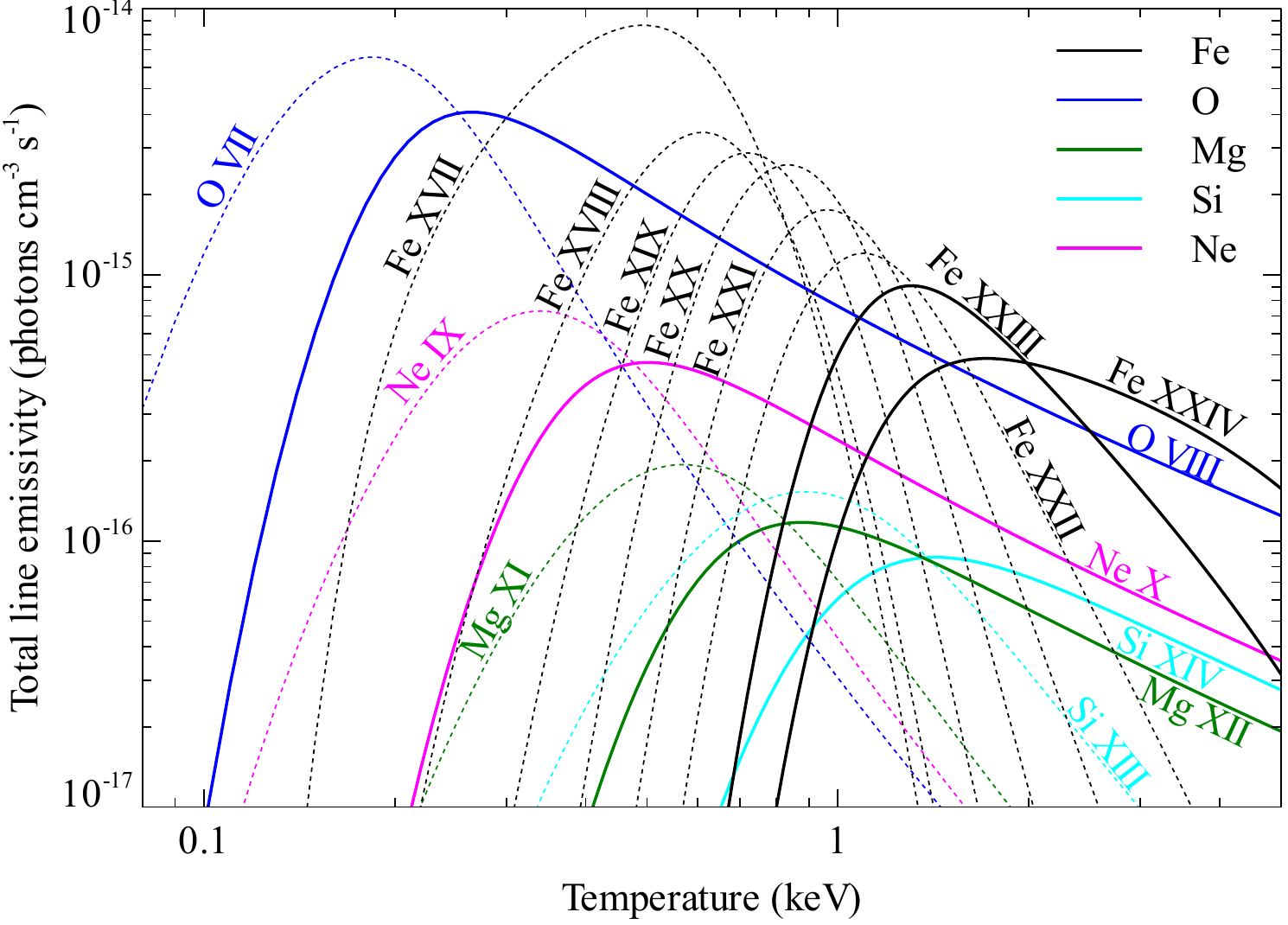}
  \caption{Total emissivity of spectral lines from different species
    as a function of temperature. The solid lines are observed in the
    RGS spectrum of A1835 and the dotted lines are not.}
  \label{fig:emissivityT}
\end{figure}

Fig.~\ref{fig:spectrum} shows the positions of emission lines that are
seen by eye in the spectrum and the positions of lines that are not
seen. All ionization states of Fe seen are Fe~\textsc{xxiii} and
above. The detections of Fe~\textsc{xxiii} are not definite as they
are blended with other lines (Fe~\textsc{xxiv} and Ne~\textsc{x}).  If
we plot the total line emissivity of ionization states seen and not
observed in the spectrum as a function of temperature
(Fig.~\ref{fig:emissivityT}), there are no obvious lines that indicate
gas at temperatures below $\sim 3$~keV. This is also seen by
comparison with the isothermal spectral models in the middle panel of
Fig.~\ref{fig:spectrum}.

If we add a further component with the same metallicities as the
first, it has a best fitting temperature of $0.58^{+0.10}_{-0.16}$~keV
and an \textsc{xspec} normalization of $6.1^{+3.4}_{-2.8}\times
10^{-5}$~cm$^{-5}$, but is only significant at the 90 per cent
confidence interval.
 
\subsection{Smoothing model (\textsc{rgsxsrc})}
As the cluster is not a point source and the RGS instruments are
slitless spectrometers, the observed spectra are broadened according
to the spatial extent of the source. This broadening is approximately
$\Delta \lambda = (0.124/m) \Delta \theta \: \textrm{\AA}$
\citep{Brinkman98}, where $m$ is the spectral order and $\Delta\theta$
is the half energy width of the source in arcmin. The effect is
relatively small in A1835, relative to nearby clusters.

Broadening was modelled using the \textsc{rgsxsrc} model in
\textsc{xspec} with an input \emph{Chandra} image in the 0.8 to 1~keV
band. The position angle of \emph{XMM} was similar between the
observations ($\sim 288^\circ$). \textsc{rgsxsrc} assumes that the
source has the same spectrum as a function of position, so is not
completely correct if different emission lines come from different
radii in the cluster.

The second column in Table \ref{tab:1t} shows the small effect of this
smoothing on the fit parameters.  The fit quality is slightly
poorer after accounting for smoothing, suggesting that the emission
lines may be emitted on a smaller scale than the broader X-ray
emission (Section \ref{sect:profiles}). The goodness command gave 54.6
per cent of realisations of the best fitting model with a better fit
than the data.

\subsection{Limits on velocity broadening of the spectra}
Since the lines are remarkably narrow, and the source extent is small,
we can limit the velocity broadening. We fitted the spectra in Section
\ref{sect:1t} with the \textsc{bvapec} model in \textsc{xspec}
assuming that the spectra were broadened by the thermal motion of the
ions in the gas (version 12.5.0ah of \textsc{xspec} fixed an error in
the line widths) and line-of-sight velocity broadening added in
quadrature. The \textsc{spex} spectral lines were used in the
\textsc{bvapec} model.

\begin{figure}
  \includegraphics[width=\columnwidth]{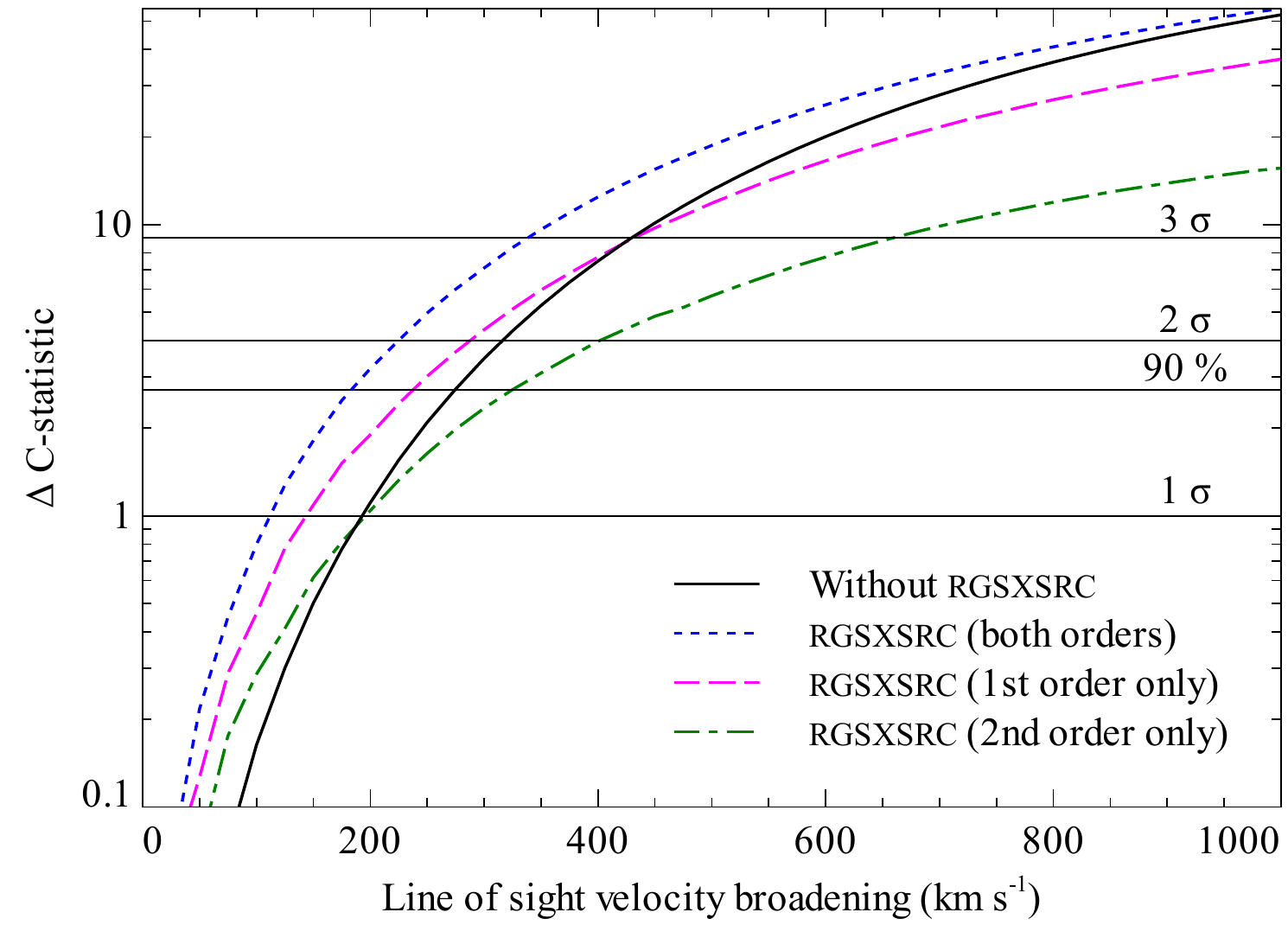}
  \caption{Limits on velocity broadening of RGS spectra. Shown are the
    increases in fit statistic as the velocity is stepped from 0 to
    $1000 \kmps$, for the data fitted with the \textsc{rgsxsrc}
    smoothing model, without the model, and then fitting the 1st and
    2nd order spectra separately.}
  \label{fig:broadening}
\end{figure}

We show the change in fit statistic as a function of velocity in
Fig.~\ref{fig:broadening}. If we conservatively treat the cluster as a
point source and do not use the \textsc{rgsxsrc} model we obtain a 90
per cent upper limit of $274\kmps$. This limit is improved to
$182\kmps$ with the addition of the broadening model. Examining just
the Fe-L spectral region between 9.6 and 12.8{\AA} (rest), we obtain a
limit of $214\kmps$, and for the 1st order spectrum between 13.6 and
22.4{\AA}, containing the strong O~\textsc{viii} line, a limit of
$380\kmps$.  We confirmed that such constraints are readily achievable
with simulated spectra.

\subsection{Limits on cooling gas}
\begin{figure}
  \includegraphics[width=\columnwidth]{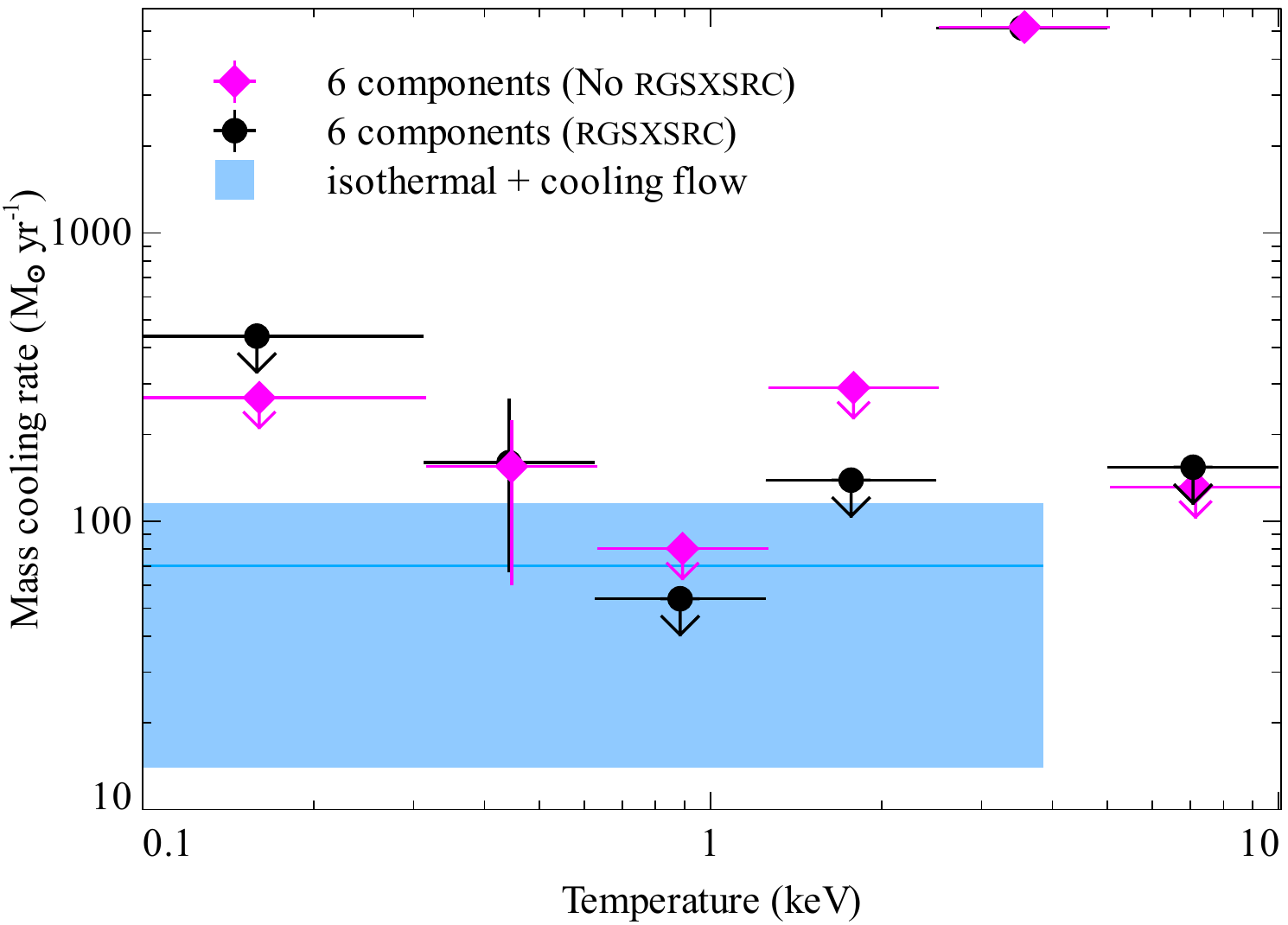}
  \caption{Mass cooling rates as a function of temperature derived
    from RGS spectra. The solid points show maximum cooling rates
    within fixed temperature bins. The shaded region shows the rate
    and uncertainty using a model made up of an isothermal plus
    cooling flow component.}
  \label{fig:coolrate}
\end{figure}

We can limit how much gas can be cooling within temperature bins. We
constructed a model using six cooling flow model \textsc{vmcflow}
components (constructed using \textsc{spex} models) with temperature
ranges of $10 \rightarrow 5.6 \rightarrow 2.8 \rightarrow 1.4
\rightarrow 0.7 \rightarrow 0.35 \rightarrow 0.0808$~keV. We fixed the
components to have the same metallicities, allowing individual
elements to vary as in Section \ref{sect:1t}. The maximum cooling
rates in $\Msunpyr$, assuming isobaric cooling, were free parameters
in the fit. The results are shown in Fig.~\ref{fig:coolrate}, with and
without the \textsc{rgsxsrc} smoothing component.

For comparison, we also fitted a model made up of an isothermal
component plus a model cooling from its temperature to zero. The best
fitting temperature of the isothermal component was 3.85~keV (close to
the single temperature fits). We obtain cooling rates from this
temperature to zero of $70^{+46}_{-56}\Msunpyr$. This rate is shown by
the shaded bar in Fig.~\ref{fig:coolrate}. The 90 per cent upper limit
is $140\Msunpyr$. A Markov Chain Monte Carlo analysis produces very
similar values. Comparing an isothermal to isothermal plus cooling
flow model with an F-test and $\chi^2$ statistics gives a chance
improvement of the $\chi^2$ of 17 per cent.

\section{Cluster property profiles}
\label{sect:profiles}
\begin{figure}
  \includegraphics[width=\columnwidth]{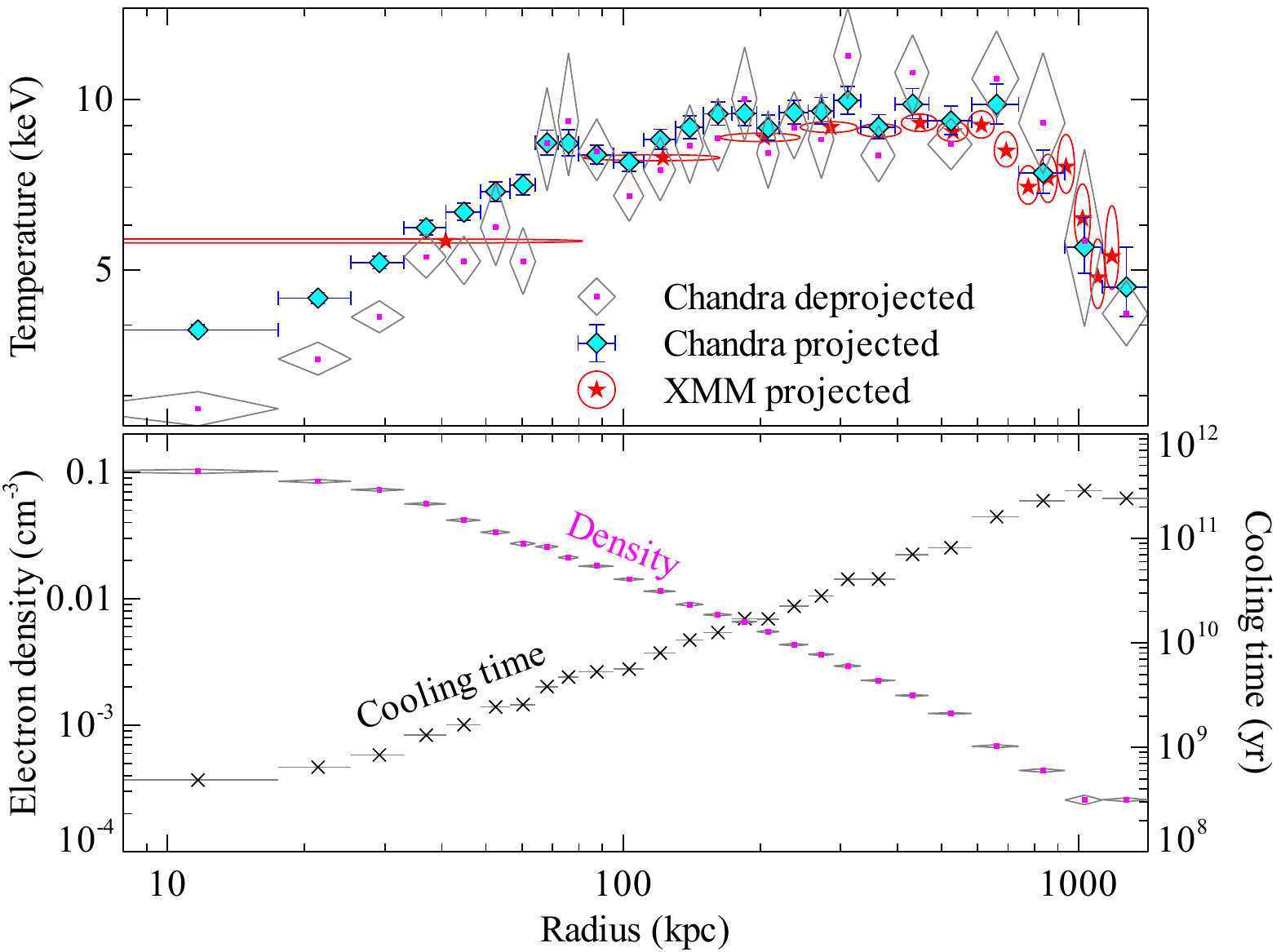}
  \caption{(Top panel) Temperature profiles, showing
      projected and deprojected \emph{Chandra} \citep{Russell08} and
      projected \emph{XMM} results. (Bottom panel) Deprojected
      \emph{Chandra} densities and cooling times. We note that there
      is a glitch in temperature at 70~kpc. The lack of density change
      associated with it is inconsistent with a shock or a cold
      front.}
  \label{fig:profiles}
\end{figure}

For comparison with the RGS results we show projected \emph{Chandra}
and \emph{XMM} EPIC-MOS temperature profiles and deprojected
\emph{Chandra} temperature, density and cooling time profiles.
(Fig.~\ref{fig:profiles}). We examined \emph{Chandra} datasets 6680,
6881 and 7370 (total 170~ks, cleaned), reprocessed applying VFAINT
filtering. Blank-sky background spectra were used, scaled to the
observed 9--12~keV count rates, providing a good match to the data in
the field outskirts.  \textsc{emchain} was used to process each MOS
dataset, removing bad time periods with a 10--15~keV rate $\ge
0.35\ps$. Scaled blank-sky spectra were used as backgrounds. There was
a small mismatch between the background and data in the outskirts
which we modelled as a uniform model component.

There is a good match between the \emph{XMM} and \emph{Chandra}
projected profiles, except where the PSF of \emph{XMM} is important in
the centre. In the core of the cluster the temperature
  drops to less than 3~keV and the cooling time is around $5 \times
  10^8\yr$. The entropy of the gas in the central region drops to $13
\keV \cmsq$. The metallicity rises smoothly inwards from 0.2 to
$0.4\Zsun$. The temperature from the RGS spectral fit, 3.7~keV, is
emitted from within 30~kpc radius (8 arcsec), essentially a point
source for the RGS.

\section{Discussion}
The limits we place on the broadening of the emission lines by
non-thermal motions of the ICM are the first direct limits on
turbulence in cool core clusters. As the cooler gas below 4~keV is
located within 8~arcsec of the core (Fig.~\ref{fig:profiles}), the
conservative point-source limit of a line-of-sight velocity of
$274\kmps$ is more appropriate than the smaller \textsc{rgsxsrc}
value. This velocity is 28 per cent of the sound speed at the measured
temperature.

Using equation 11 in \cite{Werner09} we can calculate the ratio of
turbulent to thermal energy density, $V_\mathrm{los}^2 \: \mu
m_\mathrm{p} / kT$, where $V_\mathrm{los}$ is our measured
line-of-sight velocity, $\mu$ is the mean particle mass,
$m_\mathrm{p}$ is the proton mass and $kT$ is the temperature. We
obtain a 90 per cent confidence upper limit of 13 per cent on the
energy density in turbulence.

Our upper limit is compatible with the values of 5 to 15~per cent of
pressure in gas motions, calculated using high resolution Eulerian
simulations \citep{Lau09,Vazza09}. If viscosity is
significant in clusters, as indicated by the
quasi-linear filaments in Perseus
\citep{FabianPerFilament03}, the real value is likely to be lower than
simulations predict.  The low turbulent pressure limit we measure from
the core of this cluster is encouraging for the measurement of
cosmological parameters using cluster mass profiles
(e.g. \citealt{Allen08}). Low turbulence implies that viscosity is
significant. Sound waves generated by the AGN are then damped, heating
the inner ICM \citep{FabianPer03}.

Our results also suggest that the feedback taking place in this
cluster is not inducing large amounts of turbulence in the gas in the
cluster core. The feedback must be gentle. The limit on turbulent
velocities we obtain is much lower than the turbulence required in
Perseus to prevent resonance scattering
\citep{Churazov04}. However, the region studied in
  Perseus was large and may contain differential laminar
  motions. Shearing motions, rather than turbulence, may limit
  resonant scattering. The velocity limit we find is consistent with
  the velocity shear of $250\kmps$ seen in H$\alpha$ emission for the
  central 31~kpc diameter nebula around the central cluster galaxy
  \citep{Wilman06}.

The RGS instruments have been used to measure line widths in extended
Galactic objects (e.g. \citealt{Rasmussen01}). We have shown here that
the instruments are also capable of usefully constraining
turbulent velocities for X-ray bright and peaked clusters, of which
this is the first.  Non-cool-core clusters and mergers are not so
spatially compact, making limits more difficult to
measure. \emph{ASTRO-H} will measure the ICM velocity distribution in
a wider range of objects and with much greater precision. \emph{IXO}
will map the velocity distribution over cluster cores and to higher
redshifts.

We are able to confirm the extremely low rates of gas cooling to low
temperatures first measured by \cite{Peterson01}. The distribution
does not, however, appear to be a powerlaw as seen in other objects
\citep{Peterson03}, at least below 4 keV. There is weak evidence for
an additional residual cooling component at around $100 \Msunpyr$ at
0.6~keV. This rate of cooling, and our other limits, are compatible
with an optical-spectroscopic-derived star formation rate of
40--70\Msunpyr \citep{Crawford99} and of $\sim 100 \Msunpyr$ from UV
fluxes \citep{Hicks05}, if the star formation is fed by cooling
ICM. \cite{McNamara06} suggest that the central galaxy hosts an
extremely radiatively inefficient black hole the jet power of which is
large enough to offset a large fraction of the cooling in this object.

Feedback is not completely balanced in this cluster, as is evident by
the high star formation rate, high IR luminosity and large amounts of
molecular gas. A residual level of cooling of 10 per cent of the X-ray
luminosity-inferred rate remains. Future detailed simulations and
modelling may allow us to place detailed constraints on the
temperature, velocity and spatial distribution using distant luminous
clusters like A1835.

\section{Conclusions}
For the first time we measure meaningful constraints of the ICM
turbulent velocity from the width of the X-ray emission lines. We
place a limit of 13 per cent on the turbulent energy density as a
fraction of the thermal energy density. We also constrain the
  cooling rate of X-ray emitting gas in A1835 to a value comparable to
  the star formation rate.

\section*{Acknowledgements}
ACF thanks the Royal Society for support. JRP is supported by NASA
grant \#NNX08AX45G. We thank Helen Russell for providing the
\emph{Chandra} spectra of this cluster and an anonymous referee for
helpful comments.
\bibliographystyle{mnras}
\small
\bibliography{refs}
\clearpage

\end{document}

%% file: defn.tex
\newcommand{\Mpc}{\rm\thinspace Mpc}
\newcommand{\kpc}{\rm\thinspace kpc}

\newcommand{\km}{\rm\thinspace km}

\newcommand{\cm}{\rm\thinspace cm}

\newcommand{\cmsq}{\hbox{$\cm^2\,$}}

\newcommand{\yr}{\rm\thinspace yr}

\newcommand{\s}{\rm\thinspace s}

\newcommand{\Msun}{\hbox{$\rm\thinspace M_{\odot}$}}

\newcommand{\Msunpyr}{\hbox{$\Msun\yr^{-1}\,$}}

\newcommand{\keV}{\rm\thinspace keV}

\newcommand{\erg}{\rm\thinspace erg}

\newcommand{\ergps}{\hbox{$\erg\s^{-1}\,$}}

\newcommand{\kmps}{\hbox{$\km\s^{-1}\,$}}

\newcommand{\kmpspMpc}{\hbox{$\kmps\Mpc^{-1}$}}

\newcommand{\Lsun}{\hbox{$\rm\thinspace L_{\odot}$}}

\newcommand{\Zsun}{\hbox{$\thinspace \mathrm{Z}_{\odot}$}}

\newcommand{\pcmsq}{\hbox{$\cm^{-2}\,$}}

\newcommand{\ps}{\hbox{$\s^{-1}\,$}}

